\documentclass[9pt,twocolumn,twoside]{osajnl}

\journal{ol} 

\setboolean{shortarticle}{true} 

\newcommand{\cH}{{\cal H}}

\newcommand{\rtext}[1]{\textcolor{black}{{#1}}}

\title{Rotating patterns in polariton condensates in ring-shaped potentials under bichromatic pump}

\author[1,2]{Yaroslav V. Kartashov}
\author[3,*]{Dmitry A. Zezyulin}

\affil[1]{Institute of Spectroscopy, Russian Academy of Sciences, Troitsk, Moscow, 108840, Russia}
\affil[2]{Russian Quantum Center, Skolkovo 143025, Russia}
\affil[3]{ITMO University, St. Petersburg 197101, Russia}

\affil[*]{Corresponding author: dzezyulin@itmo.ru}

\ociscodes{(190.5940) Self-action effects; (190.6135) Spatial solitons.}

\begin{abstract}
We consider a polariton condensate in a microcavity driven by a bichromatic resonant pump formed by two vortical laser beams carrying different topological charges. The system is additionally confined in a ring-shaped potential. We show that in this system steadily rotating nonlinear localized modes can be excited, whose angular rotation frequency is determined by optical frequencies and topological charges of the pump beams. When pump frequencies approach eigenfrequencies of the modes of the ring potential, resonant growth of peak amplitude of the excited  states occurs. Repulsive polariton-polariton interactions lead to tilting of the resonance curves and appearance of bistability of rotating patterns. 
\end{abstract}

\setboolean{displaycopyright}{true}

\begin{document}

\maketitle

\begin{figure*}[t!]
\begin{center}		\includegraphics[width=0.99\textwidth]{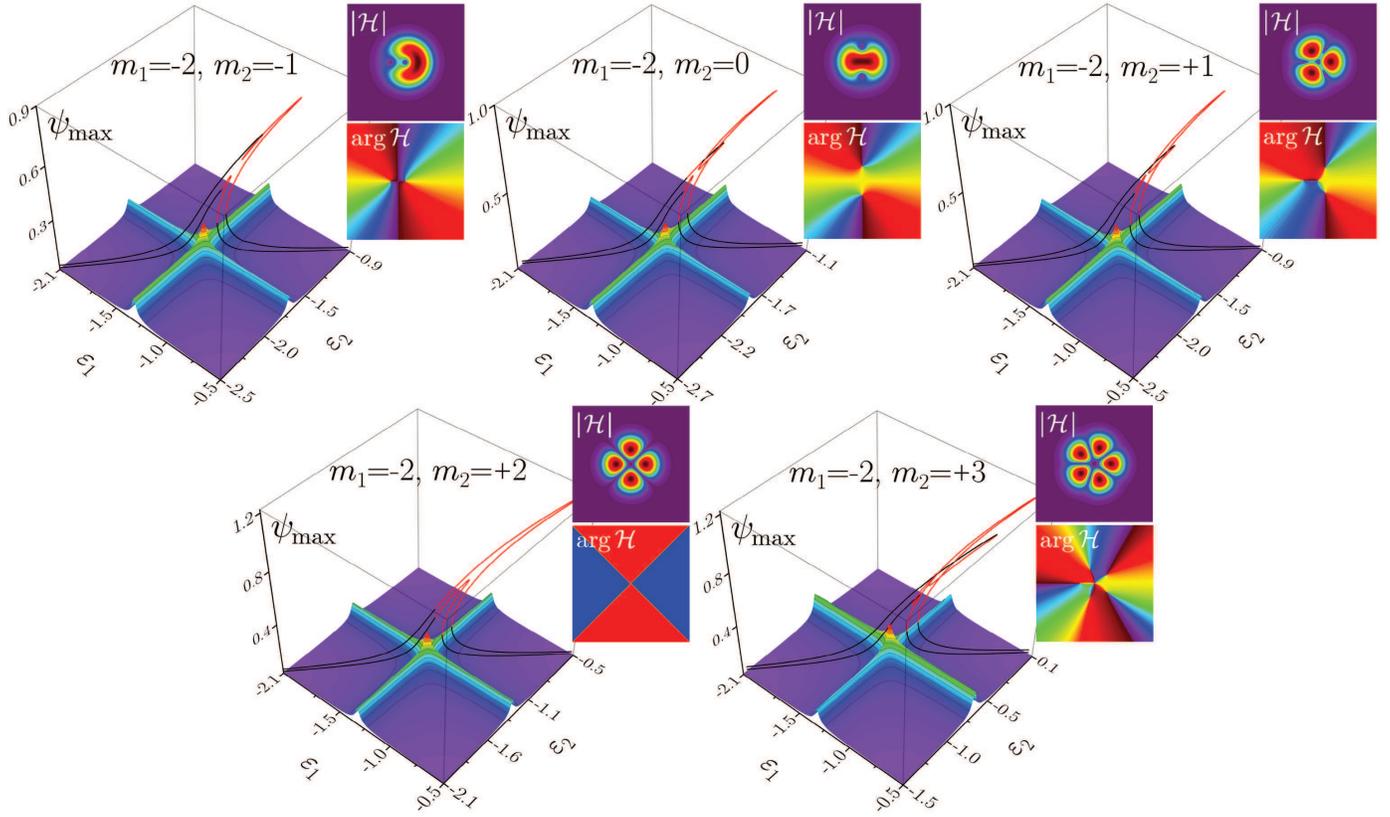}
\caption{2D surface plots illustrate linear resonance dependencies $\psi_\textrm{max}(\varepsilon_1,\varepsilon_2)$ at $h_{1,2}=0.02$  for different combinations of topological charges of the pump $m_{1,2}$. Insets show corresponding pump modulus and phase distributions at $t=0$. Lines above surface plots show nonlinear resonance curves $\psi_\textrm{max}(\delta)$  calculated along the diagonal in the $(\varepsilon_1, \varepsilon_2)$ plane passing through linear resonance point for selected topological charges $m_{1,2}$. Nonlinear resonance curves are shown for $h_{1,2}=0.05$ and $h_{1,2}=0.10$. Stable branches are shown black, unstable branches are shown red. \label{fig:fig1}}
\end{center}
\end{figure*}

Cavity polaritons are bosonic quasiparticles  that emerge in the regime of the strong coupling between the excitonic resonance and the photonic mode of the planar semiconductor microcavity \cite{Deng10}. The presence of the excitonic component results in strong polariton-polariton interactions, giving rise to various nonlinear effects and enabling Bose-Einstein condensation of polaritons. Characteristic feature of the polariton condensate is its inherently non-equilibrium nature due to photon leakage from the microcavity and recombination processes. A quasi-stationary regime can be achieved in the presence of the external pump, which can be resonant or non-resonant. In the former case, the properties of the excited patterns and excitation efficiency strongly depend on the pump shape/frequency and microcavity structuring, leading to appearance of the confining potentials.

Polariton condensates were used to demonstrate a variety of nonlinear patterns, including oblique dark \cite{dark11,dark14,resn01,nonr01,nonr04} and bright \cite{dark13} solitons. Superfluid nature of polariton condensates \cite{dark12} enables a host of associated phenomena, such as persistent currents \cite{Amo09,Sanvitto10}, Bogoliubov-like spectrum of excitations \cite{Kohnle11}, azimuthons \cite{azimuthons}, multipole solitons \cite{Zezyulin18}, polygon patterns \cite{polygons}, and nucleation of vortices \cite{Sanvitto10,dark15,Lagoudakis09,Fraser09, dark16,dark17,dark18} and vortex molecules \cite{molecules} carrying quantized topological charges. Nucleation of polariton vortices can be controlled by incoherent \cite{dark15,Dall14,Gao18,Ma18,korea} or coherent \cite{Sanvitto10,Marcetti,Gallemi18} pump. The presence of the confining potentials, which can be created using controllable metallization \cite{Lai}, etching \cite{Sala} or other techniques \cite{Schneider} of microcavity structuring, substantially facilitates the formation of excitations with desired symmetries.

Most of the works on coherently pumped polariton condensates utilize single pump beam with certain optical frequency. Formation of spontaneous extended patterns excited by two broad pump beams from the external lasers has been considered only in uniform microcavities \cite{Diaz18}. Thus, the interplay of the bichromatic pump and structuring of the microcavity remains largely unexplored.

The goal of this Letter is to describe rich resonant phenomena emerging in the presence of \textit{bichromatic} pump in the microcavity with ring-shaped potential. We consider pump beams carrying different topological charges and show that such a pump, when it is resonant with eigenmodes of ring potential, leads to excitation of steadily rotating states with nontrivial phase distributions, whose rotation frequency is determined by the optical frequencies and topological charges of the pump beams. We also discuss bistability effects appearing due to repulsive nonlinearity in polariton condensate and leading to co-existence of several stable rotating patterns. 

To model evolution of the polariton condensate in a planar microcavity under bichromatic pump, we use dimensionless Gross-Pitaevskii equation:
\begin{equation}
\label{eq:main}
i {\partial_t}\psi  = \left[-\frac{1}{2}\nabla^2 + V(r) -i\gamma + |\psi|^2 \right]\psi + \cH(x,y,t),
\end{equation}
where $\psi(x,y,t)$ is the complex-valued macroscopic wavefunction of the condensate, $\nabla^2 = \partial_{x}^2 + \partial_{y}^2$ is the Laplacian, time $t$ and spatial coordinates $x,y$.  The ring-like potential of depth $V_0$ is described by the function $V(r) = -V_0[e^{-(r-r_0)^2/w^2} + e^{-(r+r_0)^2/w^2}]$ ensuring smoothness of the potential at $r=0$, where  $r=(x^2 + y^2)^{1/2}$  is the polar radius, $r_0$ is the radius of the ring and $w$ is its width. The parameter $\gamma$ accounts for polariton losses. The function $\cH(x,y,t)$ describes bichromatic pump by two laser beams with integer topological charges $m_{1,2}$ and frequency detunings $\varepsilon_{1,2}$:
$\cH = h_1 S_{1}(r) e^{im_1\phi -i\varepsilon_1t} + h_2 S_{2}(r) e^{im_2\phi-i\varepsilon_2t}$,
where $h_{1,2}$ are the pump amplitudes, $S_{{1,2}}(r)=r^{|m_{1,2}|}e^{-r^2}$ are the functions describing \rtext{azimuthally} symmetric pump profiles, and $\phi$ is the polar angle. Equation~(\ref{eq:main}) also accounts for repulsive  interactions between polaritons.  \rtext{In order to map dimensionless Eq.~(\ref{eq:main}) to typical physical units, we assume  the  effective polariton mass   $m\approx 10^{-34}$~kg and the  unit length  $\ell=1$~$\mu$m. Then the dimensionless time unit in Eq.~(\ref{eq:main}) corresponds to $\tau=m\ell^2/\hbar\approx1$~ps. The characteristic energy $\hbar^2/m\ell^2$ corresponding to one dimensionless unit of potential depth is $\approx 0.7$~meV.}

Since bichromatic pump contains two different frequencies, the existence of steady-state modes is not obvious apriori. To show that they can exist, we move into rotating coordinate frame $x'=x\cos(\omega t) + y\sin(\omega t)$, $y'=y\cos(\omega t)-x\sin(\omega t)$, where $\omega$ is the angular rotation frequency, and where Eq. (\ref{eq:main}) reads as: 
\begin{equation}
\label{eq:rot}
\begin{array}{l}
i {\partial_t}\psi  = \left[-\frac{1}{2}(\partial_{x'}^2 +\partial_{y'}^2)  + V(r) -i\gamma + |\psi|^2 \right]\psi 
\\[2mm]+i\omega(x'\partial_{y'}-y'\partial_{x'})\psi + h_1S_{1}(r)e^{im_1\phi' -i\varepsilon_1't}+h_2S_{2}(r)e^{im_2\phi' -i\varepsilon_2't},
\end{array}
\end{equation}
where $V(r)$ and $S_{{1,2}}(r)$ do not change, since they \rtext{do not depend on polar angle}, $\phi'$ is the polar angle in the rotating frame, $\varepsilon_{1,2}' = \varepsilon_{1,2}-m_{1,2}\omega$ are pump frequency detunings in the rotating frame, and additional Coriolis term $i\omega(x'\partial_{y'}-y'\partial_{x'})\psi $ appears. Steadily rotating states of the form $\psi = u(x',y')e^{-i\mu t}$ are possible in Eq.~(\ref{eq:rot}) if all quantities evolve in time with the same rate $\varepsilon_1'=\varepsilon_2'=\mu$, where $\mu$ has the meaning of the effective energy. This   determines the rotation frequency of the pattern:
\begin{equation}
    \label{eq:omega}
    \omega = {(\varepsilon_1 - \varepsilon_2)}/{(m_1-m_2)}.
\end{equation}
that depends on both detunings $\varepsilon_{1,2}$ and topological charges $m_{1,2}$ of the pump, as well as effective energy: 
\begin{equation}
    \label{eq:mu}
    \mu = {(m_1\varepsilon_2 - m_2\varepsilon_1)}/{(m_1-m_2)}.
\end{equation}
The most representative feature of the resonant pump is that the amplitude of the modes that it excites resonantly increases when frequency of the pump approaches eigenfrequencies of corresponding linear eigenmodes of the system. Moreover, efficient excitation of such modes requires matching of their topological charge with topological charge of the pump. Our ring-like potential $V(r)$ supports finite number of localized modes with different topological charges that can be found from eigenvalue problem $\mu_{\rtext{m}} u_m  = [-(1/2) \nabla^2  + V(r)]u_m$ obtained from linear version of Eq.~(\ref{eq:main}) at $\gamma=0$, where $\mu_m$ and $u_m$ are the eigenfrequency and eigenmode profile corresponding to topological charge $m$. For representative depth $V_0=6$, width $w=0.25$, and radius $r_0=2$ that we use here, the ring-like potential supports four modes with eigenfrequencies $\mu_0=-1.865$, $\mu_{\pm1}=-1.719$, $\mu_{\pm2}=-1.312$, and $\mu_{\pm3}=-0.676$, three of which are degenerate, because they correspond to opposite values of topological charge $m$. When pumping simultaneously at two frequencies in \textit{linear system} one excites combinations of two modes with topological charges dictated by the pump, whose interference results in patterns rotating with the frequency (\ref{eq:omega}). The weight of each mode in superposition depends on how close is pump frequency $\varepsilon_{1,2}$ to the eigenfrequency of this mode. The dependence of the maximal amplitude \rtext{$\psi_\textrm{max}=\textrm{max}|\psi|$} of resulting rotating pattern on $\varepsilon_1$ and $\varepsilon_2$ in the linear case is shown in the form of 2D surfaces in Fig.~\ref{fig:fig1} for five different combinations of the topological charges in the pump components (further we set $m_1=-2$ and gradually increase $m_2$) and equal pump amplitudes $h_1=h_2$.
Each surface involves two perpendicular crests, which emerge at $\varepsilon_1$ and $\varepsilon_2$ values coinciding with eigenfrequencies $\mu_{m_1}$ and $\mu_{m_2}$ of the modes of the potential. Their intersection produces a spike, where two modes interfere with maximal amplitudes, that corresponds to the case when both pump frequencies are in resonance with eigenmodes of the potential. Like for any driven dissipative system, the height of the spike decreases with $\gamma$ (hereafter  we use typical for polaritons value of losses $\gamma=0.02$).

\begin{figure}[t!]
\begin{center}		\includegraphics[width=\columnwidth]{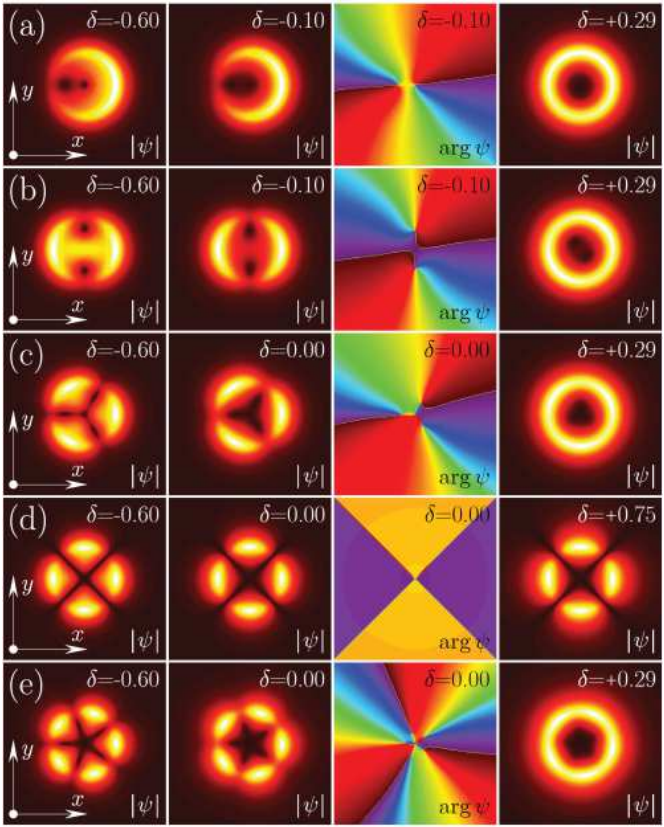}
\caption{ Examples of modulus distributions of polariton wavefunction $\psi$ at $h_{1,2}=0.1$ for different values of detuning $\delta$ and different combinations of topological charges of the pump components: (a) $m_2=-1$, (b) $m_2=0$, (c) $m_2=+1$, (d) $m_2=+2$, (e) $m_2=+3$, while $m_1=-2$ in all cases. All states are taken from the ``upper'' branch that smoothly continues to large negative  $\delta$ values. Third column shows phase distributions $\arg\psi$ corresponding to $|\psi|$ distribution from the second column. All plots are shown in  $x,y\in [-5, 5]$   window.\label{fig:profiles}}
\end{center}
\end{figure}

\begin{figure}
\begin{center}		\includegraphics[width=\columnwidth]{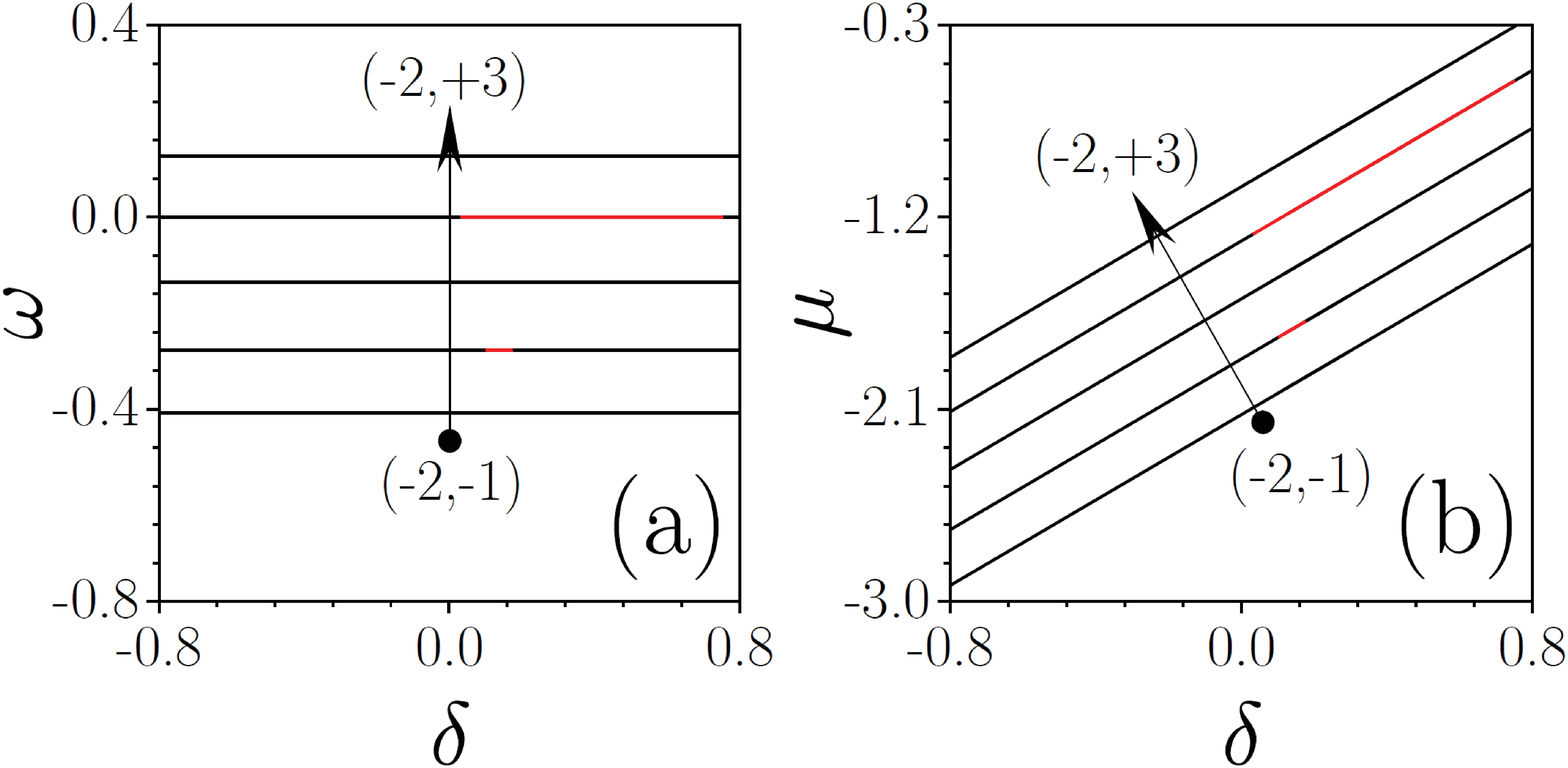}
\caption{Effective rotation frequency (a) and energy (b) \textit{vs.}  detuning $\delta$ for diagonal scan on the $(\varepsilon_1, \varepsilon_2)$ plane and for different combinations of topological charges $(m_1, m_2)$ in the pump. In all cases $m_1=-2$,  while direction of increase of $m_2$ from -1 to +3 is shown by arrows. Red segments correspond to detuning intervals, where   upper branch is unstable at $h_{1,2}=0.10$. \label{fig:lines}}
\end{center}
\end{figure}

In the presence of nonlinearity resonances become asymmetric: they tilt progressively toward larger $\varepsilon_{1,2}$ values with increase of pump amplitude $h_{1,2}$, so that at one point bistability emerges. To demonstrate this effect we scan the parameter plane $(\varepsilon_1, \varepsilon_2)$ along the diagonal passing through linear resonance at $\varepsilon_1=\mu_{m_1}$ and $\varepsilon_2=\mu_{m_2}$. We thus introduce and change the parameter $\delta$ --- frequency detuning from linear resonance --- such that pump frequencies vary as $\varepsilon_1(\delta) = \mu_{m_1} + \delta$ and $\varepsilon_2(\delta) = \mu_{m_2} + \delta$. Calculated \textit{nonlinear} resonance curves $\psi_\textrm{max}(\delta)$ for two different amplitudes of the bichromatic pump are shown with solid lines on top of 2D surfaces in Fig.~\ref{fig:fig1}. Progressively increasing nonlinearity-induced tilt of resonances leading to coexistence of several solutions is obvious. For $h_{1,2}=0.10$ the upper branch (that can be smoothly continued to large negative $\delta$ values) tends to create a loop, that leads to coexistence of five states, for $h_{1,2}=0.05$ up to three states can coexist for the same $\delta$. The symmetry of corresponding rotating states is determined by the topological charges of the pump, see Fig.~\ref{fig:profiles}. For negative and zero detuning $\delta$ all shown patterns from the upper branch feature distinctive multipole shapes with nontrivial phase distributions with several nested vortices (see third column of Fig.~\ref{fig:profiles}) and the number of "petals" equal to $|m_1-m_2|$, in agreement with pump modulus distribution $|\mathcal{H}|$ (see insets in Fig.~\ref{fig:fig1}). However, with increase of $\delta$ to sufficiently large positive values in all such states nested vortices move to the center and the pattern becomes nearly ring-like (see fourth column of Fig.~\ref{fig:profiles}), with the only exception for state generated by $m_1=-2,~m_2=+2$ pump that does not rotate for detuning values taken from the diagonal on the $(\varepsilon_1,\varepsilon_2)$ plane \rtext{(if one moves along any other line in this plane this state also transforms into ring-like structure)}. Substituting $\varepsilon_1 = \mu_{m_1} + \delta$ and $\varepsilon_2 = \mu_{m_2} + \delta$ into Eqs.~(\ref{eq:omega}) and (\ref{eq:mu}), one obtains the following expressions for rotation frequency and and energy:
     $\omega = {(\mu_{m_1} - \mu_{m_2})}/{(m_1-m_2)}$, $\mu= \delta + {(m_1\mu_{m_2} - m_2\mu_{m_1})}/{(m_1-m_2)}$,
i.e., when we move along the diagonal in the $(\varepsilon_1,\varepsilon_2)$ plane, the rotation frequency of the pattern does not change, while energy changes linearly with $\delta$. These simple observations are illustrated in Fig.~\ref{fig:lines}.

\begin{figure}[t!]
\begin{center}		\includegraphics[width=\columnwidth]{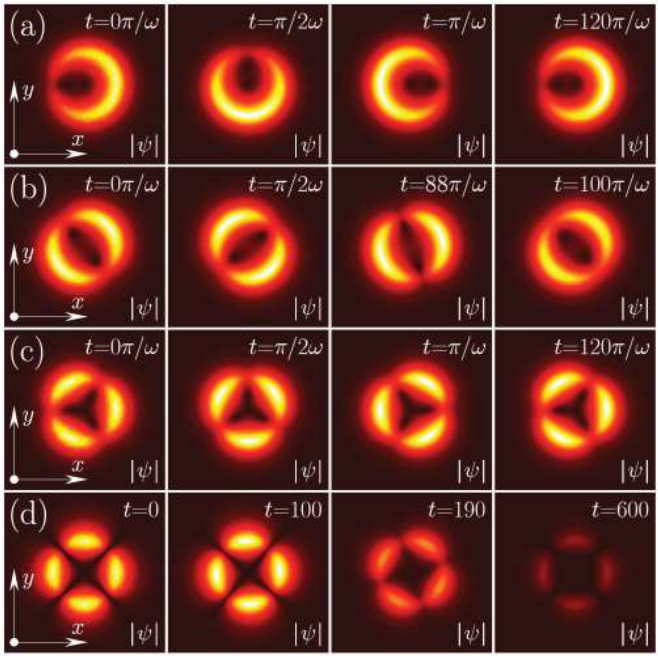}
\caption{Examples of stable evolution of the rotating states at $\delta=-0.1$, $m_2=-1$ (a) and $\delta=0$, $m_2=+1$ (c). (b) Instability development at $\delta=+0.1$, $m_2=0$ leading to formation of  persistent breather  with the period $\approx 22.6\pi/\omega$. (d) Instability development leading to switching of the state from the upper to lower branch at $\delta=+0.29$, $m_2=+2$. In (a),(c), and (d) pump amplitude $h_{1,2} =0.10$, while in (b)
$h_{1,2} =0.05$. All input states are taken from the upper branch that smoothly continues to negative $\delta$ values. All distributions are shown within $x,y\in [-5, 5]$ 
window. See \textcolor{blue}{Visualizations 1,2,3,4}. \label{fig:dyn}}
\end{center}
\end{figure}

Stability of the rotating states 
is determined by the pump amplitude $h_{1,2}$ and detuning $\delta$. To test stability, 
we have employed the standard approach based on the linearization \rtext{\cite{Yang}} of  Eq.~(\ref{eq:main}) with respect to small perturbation of a steadily rotating state. The outcome of the linear stability analysis have been also confirmed by the direct solution of Eq.~(\ref{eq:main}) with initial conditions corresponding to slightly perturbed rotating states. The results of the stability analysis are indicated on the nonlinear resonance curves in Fig.~\ref{fig:fig1}, where black and red segments correspond, respectively, to stable and unstable states. 
Lower branches are always stable, upper branches are usually stable up to the critical value of $\delta$ (typically corresponding to the point, where this branch starts to make a loop), and middle branches are unstable. Examples of persistent rotation of perturbed modes from stable portions of the upper branch are presented in Figs.~\ref{fig:dyn}(a,c). These modes do not change their symmetry over multiple rotation cycles. Unstable modes show different scenarios of instability development. Close to the critical value of $\delta$, where instability emerges, one observes formation of persistent breathers (states with oscillating peak amplitude) with periods far exceeding rotation period [see Fig.~\ref{fig:dyn}(b) where reorientation of the pattern, accompanying amplitude breathing, is obvious from last two columns]. Unstable states close to the tip of the resonance curve usually switch to small-amplitude stable state from the lower branch, \rtext{sometimes after slight initial reorientation,} see Fig.~\ref{fig:dyn}(d). 
Stable rotating patterns are attractors that can emerge from the input having very distinct shape --- in the course of evolution they start rotating with the frequency given by Eq.~(\ref{eq:omega}).

To conclude, we have demonstrated that polariton microcavity with ring-like potential in the presence of bichromatic resonant pump supports a variety of rotating stable states, whose spatial shape and rotation frequency are fully determined by the topological charges and frequencies of the pump beams.  Experimental relevance of our findings is confirmed by the fact that rotating vortices driven by pulsed resonant excitation have been already observed in polariton condensates in the noninteracting regime \cite{new}.

\section*{Funding Information} Russian Science Foundation (Project 17-12-01413).



\newpage

\section*{Full references with titles}
  

\ifthenelse{\equal{\journalref}{aop}}{%
\section*{Author Biographies}
\begingroup
\setlength\intextsep{0pt}
\begin{minipage}[t][6.3cm][t]{1.0\textwidth} 
  \begin{wrapfigure}{L}{0.25\textwidth}
    \includegraphics[width=0.25\textwidth]{john_smith.eps}
  \end{wrapfigure}
  \noindent
  {\bfseries John Smith} received his BSc (Mathematics) in 2000 from The University of Maryland. His research interests include lasers and optics.
\end{minipage}
\begin{minipage}{1.0\textwidth}
  \begin{wrapfigure}{L}{0.25\textwidth}
    \includegraphics[width=0.25\textwidth]{alice_smith.eps}
  \end{wrapfigure}
  \noindent
  {\bfseries Alice Smith} also received her BSc (Mathematics) in 2000 from The University of Maryland. Her research interests also include lasers and optics.
\end{minipage}
\endgroup
}{}

\end{document}